\documentclass[amsmath,amssymb,nofootinbib,prd]{revtex4}

\usepackage{setspace}
\usepackage{graphicx}

\usepackage{amsmath,amssymb,amsfonts,amsthm}
\usepackage{enumerate}

\begin{document}
\title{Quantum panprotopsychism and the structure and subject-summing combination problem}

\author{Rodolfo Gambini$^1$,  Jorge Pullin\footnote{Corresponding author. Email: pullin@lsu.edu}$^2$}
\affiliation{1. Instituto de F\'{\i}sica, Facultad de Ciencias, Igu\'a 4225, esq. Mataojo,
11400 Montevideo, Uruguay. \\
2. Department of Physics and Astronomy, Louisiana State University,
Baton Rouge, LA 70803-4001, USA.}

\begin{abstract}
In a previous paper, we have shown that an ontology of quantum mechanics in terms of states and events with internal phenomenal aspects, that is, a form of panprotopsychism, is well suited to explaining the phenomenal aspects of consciousness. We have proved there that the palette and grain combination problems of panpsychism and panprotopsychism arise from implicit hypotheses based on classical physics about supervenience that are inappropriate at the quantum level, where an exponential number of emergent properties and states arise.  
In this article, we address what is probably the first and most important combination problem of panpsychism: the subject-summing problem originally posed by William James. We begin by identifying the physical counterparts of the subjects of experience within the quantum panprotopsychic approach presented in that article.  To achieve this, we turn to the notion of subject of experience inspired by the idea of prehension proposed by Whitehead and show that this notion can be adapted to the quantum ontology of objects and events. Due to the indeterminacy of quantum mechanics and its causal openness, this ontology also seems to be suitable for the analysis of the remaining aspects of the structure combination problem, which shows how the structuration of consciousness could have evolved from primitive animals to humans.  The analysis imposes conditions on possible implementations of quantum cognition mechanisms in the brain and suggests new problems and strategies to address them. In particular, with regard to the structuring of experiences in animals with different degrees of evolutionary development.
\end{abstract}
\maketitle
\section{Introduction}

Consciousness has recently been subject to a huge number of different approaches. Although we have direct access to conscious experience, there is nothing harder to explain. A theory of consciousness should at least identify the physical processes that give rise to consciousness. Although we would like to understand consciousness as an integral part of the natural world, purely physicalist explanations fall short of explaining the phenomenal, purely qualitative aspects of consciousness. Purely reductive, naturalistic methods to explain consciousness end up treating qualia as illusory. In fact, naturalism maintains in its various versions that reality is completely analyzable in scientific terms. Things that do not appear to admit a scientific description, such as qualia, normativity, responsibility, or freedom, must be described by science or explained away by science \cite{decaro}. Without consciousness, there is no personal life. Chalmers identifies the problem of experience, understanding the subjective qualitative aspect, as the ``hard problem of consciousness" \cite{chalmersfacing} p.203. He has analyzed the possibility of explaining it resorting to the notion of natural supervenience assuming the existence of new natural laws that connect the physical with the phenomenal. It is therefore a form of property dualism. However, since he implicitly assumes the causal closure typical of classical physics, the new {\em psychophysical laws} connecting the physical with the phenomenal cannot interfere with the physical processes. These laws of supervenience together with the supposed causal closure of the physical cause the phenomenal experience to behave as an epiphenomenon. This contradicts our own ability to reflect on the existence and meaning of conscious phenomena and, as we will see, falls short when it comes to explaining the development of our perceptual abilities.

An alternative to this view is regularism, the notion \cite{consciousness3} that the regularities that physics identifies in its description of processes and phenomena are fulfilled without exception, without asserting that nature is interchangeable or exhausted by the physical description. This is consistent with the structuralist conception of science shared by many philosophers and scientists like Wittgenstein, van Fraassen, or Hawking that distinguishes between the mathematical laws and the world that they describe. As regularism does not assume that the world is exhausted by the physical, a quantum probabilistic description is open to novel causal powers such as the ones resulting from the phenomenal realm. 

If the phenomenal world has a degree of independence from the physical world and does not supervene on it, panpsychism is its most economic description. Any object whose third-person empirical description is physical could have a phenomenal intrinsic nature. Usually, in panpsychism, a kind of mind-like element is attributed to elementary particles like quarks and photons. As we shall see, quantum physics teaches us that this kind of attribution is incorrect, the fundamental concepts of quantum mechanics according to its axioms are systems in certain states that produce events in their interaction with other systems. This ontology of states and events should be the basis of a panprotopsychic vision resulting from quantum mechanics. We have analyzed this possibility in a previous paper showing that it allows to solve several aspects of the combination problem \cite{consciousness3}, such as the palette and grain problems. Here we shall extend this analysis to two of the oldest and most important objections to panpsychism: the structural and subject-summing combination problems related to the nature of subjectivity, in particular of proto-subjectivity. 
In Section II we discuss the hard problem of consciousness and analyze the limitations imposed by a physicalist treatment of it. In Section III, we introduce the notion of regularism as an attempt to explore the limits imposed by the physical description of the material world. In Section IV we analyze the type of panpsychism supported by a quantum description of the world. We consider in detail what the phenomenal aspect of elementary physical processes may be, exploring the notion of quantum panprotopsychism. In Section V, we apply this notion to the different subproblems of the combination problem of panpsychism with special emphasis on the structure and subject-summing combination problem. Finally, in Section VI, we summarize some conclusions and open problems.

\section{Physics and the hard problem of consciousness}

Consciousness provides the only direct way to access the world. We all have conscious experiences that manifest in various forms in sounds, colors, smells, or feelings of pain or joy. They all have qualitative, personal, and noncommunicable manifestations that are generally called phenomenal qualities or qualia. Daniel Dennet considers that ``human consciousness is just about the last mystery to survive. ... Consciousness stands alone today as a topic that often leaves even the most sophisticated thinkers tongue-tied and confused." It is such a central element in our lives that we consider a life without consciousness not worth living. Without consciousness, there is no personal life. Chalmers distinguishes between ``the easy problems of consciousness [which] are those that seem directly susceptible to the standard methods of cognitive science, whereby a phenomenon is explained in terms of computational or neural mechanisms'\cite{chalmersfacing} p.202 and the hard problems that seem to resist those methods. ``The really hard problem of consciousness is the problem of experience. When we think and perceive, there is a whir of information processing, but there is also a subjective aspect. There is something like being a cognitive agent. When we see, for example, we experience visual sensations: the felt quality of redness, the experience of dark and light, the quality of depth in a visual field." \cite{chalmersfacing} p.203.

\subsection{Phenomenology}
We access consciousness directly and through it, everything else we know. However, in our daily lives we experience the physical or cultural objects around us, as well as people, as the basic elements of the world in which we live. The primitive objects that constitute our daily reality, whose existence cannot and should not be questioned. This ``natural attitude,” in Husserlian terms, characterizes our habitual way of being in the world. Generally, we are so absorbed in the facts, in our objectives, desires, and fears that the essential functions of our mind are barely perceived and remain implicit. The enormous development of our scientific understanding of the world has led to the paradoxical situation that consciousness, that is, that to which we have the most direct and intimate access, is the least understood. We are torn between considering consciousness as a mystery and taking it as an illusion.

Our conscious experiences have an internal aspect. We all know what it is like to experience pain, pleasure, a soft aroma of coffee, or the sounds of a piece of music. In all cases, we experience phenomenal qualities or qualia. Explaining how these conscious qualities emerge and what their relationship is with the world described by the sciences, in particular physics, is what Chalmers \cite{chalmersfacing} calls the hard part of the mind-body problem. More precisely, the problem is to include in a coherent description the third-person world described by our natural sciences and the first-person world of our conscious experiences. This does not necessarily imply the reduction of one to the other.

Phenomenology is an important tool for the study of the first-person perspective. Husserl \cite{ideas} developed phenomenology to investigate the features of consciousness avoiding assumptions about the external world and exploring the meaning and significance of our experiences. For Husserl, following Brentano \cite{brentano}, an essential aspect of consciousness is its intentionality. In general terms, being conscious means being aware of something. Every act of consciousness is an attribution of meaning, it is the consciousness of something. That process of directing oneself towards something and attributing it a meaning, a noema, is what Husserl calls intentionality. It is our most common form of consciousness and accompanies our basic actions of looking, hearing, or touching, but also of imagining, remembering, or desiring. Awareness of any real or imaginary object involves an intentional approximation to it \cite{consciousness3}.

Husserl \cite{ideas} refers to intentionality as a complex of relations that includes us as subjects, our perceptive acts, their conceptualization in what he calls noema, and the natural or cultural object that generate them. Intentionality is only the first manifestation of a consciousness that reflects on itself \cite{consciousness3}. But the task of phenomenology is more general; it consists in reflecting on the phenomena that occur in our consciousness, leaving aside their physical nature or their cultural origin. It provides a rigorous first-person tool for the analysis of the phenomenal aspects of consciousness and recognizes its effectiveness beyond its pure qualitative aspects. It also allows us to glimpse the elemental characteristics of all conscious phenomena, including those that occur in other people and very possibly in many other living beings. 

But ``The essence of transcendental idealism for Husserl was the a priori correlation between objectivity and subjectivity" \cite{dictionary} p.330. ``The world is the correlate of consciousness" \cite{ideas} p.147. Consciousness is the primary source of all evidence, but the transcendental idealism of Husserl, based on the notion of intentionality, is directed toward a reality that hides behind the world that we intentionally inhabit, of which we can only have a hint coming from direct contact with ourselves \cite{consciousness3}.

\subsection{The physical--mathematical description of reality}

The progress of physics originates in the mathematical representation of phenomena. In a broad sense, a representation consists of the use of one thing to select and highlight properties of another. Van Fraassen \cite{vfr} has observed that every representation has intentionality in Brentano's sense: ``Representation is intentional in the sense of being about something in just the way that reference... and predication are.” Thus, we can represent the evolution of Covid cases during the 2020-2021 pandemic using a mathematical function. Every form of representation involves a selection of certain aspects of the phenomenon and an abstraction. It looks for some type of similarity between the representative object and the represented one. Mathematical descriptions of phenomena, which in many cases involve measurements, play an essential role as a starting point of the mathematical representation process. Modern mathematical physics started from the Copernican observation that a heliocentric system seen from Earth would present the same appearances as the Ptolemaic system. They are two theoretical descriptions physically associated with different reference systems that can be considered as two representations that approximate the same set of appearances and lead approximately to the same set of measurements. The Copernican vision, in addition to its elegance and simplicity, was accepted for its fruitfulness. Kepler's laws and the establishment of the principles of modern physics with Galileo, mainly his law of inertia and his principle of relativity, are direct consequences of the adoption of the heliocentric description of the Solar System. 

It was Galileo's genius who discovered that mathematics is the grammar of science. It is this discovery of the rational structure of nature that gave the a priori foundations of modern experimental sciences and made its constitution possible \cite{koyre}. A successful representation leads to overvaluing the well-represented aspects of the phenomenon. The introduction by Galileo of its primary qualities as the only real ones and the elimination of the secondary ones as illusory is an example of this overvaluation of what is directly mathematically representable and accessible by our different senses. In this way, Galileo adopted an ontological thesis assuming that his mathematical models describe a reality which is mathematical in nature. For that, he needs to assume that secondary qualities would not exist at all without human sensory organs to produce these qualitative features.

At the end of the 19th century, with Hertz and Poincare, we began to have a more critical vision of what is described by scientific theory \cite{vfr} p.196. Hertz says ``We form inner pictures or symbols of external objects; and the form which we give them is such that the necessary consequences of the images in thought are always the pictures of the necessary consequences of the things pictured." What needs to be elucidated is how we form those inner images. The previous discussion of representation suggests the answer.  We need to connect the physical description to the world to which it refers. This connection is made by attributing mathematical properties to certain observed objects through measurements. 

There are entities that even defenders of empirical structuralism such as van Fraassen admit as real, they are the correlates of observable phenomena. He wonders \cite{vfr} “Is there something that I could know to be the case, and which is not expressed by a proposition that could be part of some scientific theory?”, and answers “YES something expressed only by an indexical proposition” that is, a phenomenon. These phenomena do not have to be mathematical in nature, but we can always represent them by attributing mathematical properties to them. In the case of the planets, an example would be their location in the celestial sphere as seen from Earth. From the phenomena, an empirical description is constructed, such as Kepler's laws, from carefully compiled astronomical data. The theoretical work that remains to be done is to obtain the empirical structure given in Kepler's laws from a model of the Solar System. The matching that Newton demonstrated was therefore between mathematical structures, by recovering the empirical description from his three fundamental laws, and his law of universal gravitation. The objective of any physical description is the recovery of the mathematical structures obtained through measurements of the phenomena associated with indexical propositions of a certain region of interest. At no point are the structures supposed to require any kind of similarity between the manifest and the scientific \cite{consciousness3}.
As emphasized in Husserl's ``Crisis'' \cite{crisis2}, with Galileo, mathematicality becomes a criterion for existence \cite{consciousness3}. In his later works, Husserl introduced the concept of the lifeworld, which recognizes that even at its deepest level, consciousness is already operating in a world of meanings and prejudices that are socially, culturally, and historically constituted through an intersubjective process that begins at birth. It involves everything from the implicit foundations of natural science to the set of assumptions with which we conduct our daily lives. What is given or assumed immediately in our behavior. Without a set of assumptions, no experience is possible. For Husserl, the scientific vision of the world must be understood from this lifeworld of our experiences that has been forgotten by our scientific description.

Husserl considers that the scientific vision that originated with Galileo and persists in many scientists is based on a set of assumptions. These can be characterized by the surreptitious substitution of the mathematically subtracted world of idealities for the only real world, the one that is actually given through perception , that is ever experienced and experienceable \cite{Crisis} leading in this way to the implicit substitution of the objects of the world by mathematical idealities. Abstracting from the human beings their cultural and spiritual dimensions and leading therefore \cite{celestino} to problems concerning the status of the human sciences. This process of mathematization of reality that began with Galileo has not only not been overcome but some even today propose to deepen it. Those who support this point of view in most cases when analyzing the phenomenon of consciousness tend to reduce it to some form of awareness that allows us to access certain information from the environment or our body and use it to control behavior, forgetting the phenomenal side of conscious experience.
Tegmark \cite{tegmark} is perhaps the one who has made more explicit that position. He maintains that \cite{tegmark} p. 104: ``Our external physical reality is a mathematical structure..." He distinguishes between ``the outside view or bird perspective of a mathematician studying the mathematical structure and the inside view or frog perspective of an observer living in it." Tegmark's Mathematical Universe Hypothesis, like Galileo's original position, is a form of mathematical monism where reality is reduced to the mathematical structure. Everything else, the phenomenal consciousness and the qualitative aspects of experience, is relegated to the mere status of an illusion, deemed nonexistent \cite{consciousness3}. This position seems to forget the very origin of the process of abstraction of the sciences which, as we have highlighted, projects the phenomenal into the mathematical through observation and measurement. It also considers secondary qualities as irrelevant ---as ``mere baggage"---. This position was reasonable in Galileo's times because the new physics said nothing about them. But today we know that secondary qualities such as color, smell, or sound are associated with well-understood physical events, so this bifurcation of reality does not seem justifiable. The physical aspects of the secondary qualities taken by Galileo as a mere illusion were, over time and the development of physics, mainly quantum mechanics, incorporated into the mathematical description. As Locke observed: It is one thing to perceive and know the idea of white or black, and quite another to examine what kind of surface texture is needed to make an object appear white or black. More explicitly, he establishes that \cite{locke4} Book 2 Chap.8 ''Whatsoever the mind perceives in itself, or is the immediate object of perception, thought, or understanding, that I call IDEA; and the power to produce any idea in our mind, I call QUALITY of the subject wherein that power is. Thus a snowball having the power to produce in us the ideas of white, cold, and round,—the power to produce those ideas in us, as they are in the snowball, I call qualities; and as they are sensations or perceptions in our understandings, I call them ideas''  It is those qualities of things that are explained by modern physics in terms of the production of events in those things. How they produce our perceptions ---the Lockean ideas--- through our senses is part of the hard problem of consciousness.

The distinction between primary and secondary qualities is totally arbitrary and only obeys to the degree of development of physics in Galileo's times. Physics explains both the emission of radiation of certain wavelengths by a metal corresponding to a certain color and the motions of massive particles in a certain space-time geometry in terms of the same basic concepts: {\em states and events}. In fact, metals are colored because the events corresponding to the absorption and reemission of light depend on the wavelength. For example, copper has low reflectivity at short wavelengths and preferentially emits red to yellow frequencies. Using the Galilean terminology, both phenomena admit mathematical descriptions in terms of events, the first linked to space, time, and moving particles would be associated with really existing primary qualities, while the second with secondary qualities coming from our senses. Considering the former as real and the latter as illusory is today unjustifiable.  Tegmark opts for a second option, only the physical-mathematical description of both is real and the perceptions that they awaken in us ---the Lockean ideas--- would be illusory in both cases. Faced with these two interpretations, a third alternative arises which, as shown, by van Fraassen and proposed by Hertz, Mach, Poincaré and others, considers the physical description as the result of a process of abstraction and idealization of an external reality, to which we can only refer indexically. The apparent difference between the qualities associated with the spatio-temporal properties that Galileo treats as primary and the others is that our perceptual system organizes sensations coming from different senses, attributing them a coherent spatial distribution, but both originate in events. We will return to this point later. As we shall see, quantum mechanics allows us to identify the basic elements of the external reality as systems in certain {\em states} that manifest themselves ostensibly in phenomena or {\em events}. 

All fundamental physical theories, whether classical like general relativity, or quantum mechanical, have elements in common. They have a mathematical framework, an interpretation, and correspondence rules that connect the mathematical formalism to phenomena. They provide mathematical descriptions of certain physical systems. Initially, the system will be in a certain {\em state} and will evolve giving rise to {\em events} that constitute the observable phenomena.  To them ---the states and events--- we can associate certain mathematical objects. The mathematical regularities that will result from the description provided by the theory will be fulfilled without exception, although in the case of quantum theories, they will have a probabilistic nature \cite{persons}.

Events took a central role with the advent of quantum mechanics at the beginning of the twentieth century \cite{consciousness3}. The new theory extends the description of nature to a new sphere of reality: the microscopic. It includes atomic and molecular phenomena, among others. Biological life at the molecular level depends strongly on the quantum properties of certain large molecules called proteins. Since large objects are composed of atoms or molecules, quantum mechanics is the most fundamental theory, its effects transcend the microscopic world, and they are necessary, for instance, for understanding the physical properties of solid and liquid materials. Our current understanding is that classical physics is only one approach that is applicable to the study of some macroscopic properties of quantum systems that have lost their coherence, which is common in many systems at room temperature. However, the world, as described in the third person, is fundamentally quantum in nature. To understand the properties of the objects that surround us, one needs to resort to quantum theory. For example, if we wish to determine the electrical or thermal conductivity or the color that we will observe in a copper cable, we must resort to quantum theory \cite{persons}. The same happens in many biological systems that present behaviors that depend on quantum mechanics such as tunneling, superposition, coherence, and entanglement \cite{mcfadden}, \cite{iw}. Adams and Petruccione \cite{ap} recently published a review that analyzes several possible quantum effects involved in the brain: ``the firing of nerves; the actions of anesthesia, neurotransmitters, and other drugs; the sensory interpretation and organized signaling that are central to the vast neural network that we identify as our self". Particularly interesting is Fisher's model of quantum cognition \cite{fisher}, which is a possible implementation of the quantum phenomena that we will discuss here \cite{consciousness3} and involve entangled Posner molecules. Another mechanism of entanglement, in this case of photons in the brain, was recently discovered \cite{liu}.

Thus, we consider states and events as the fundamental categories in a theory that correlate to what exists in a quantum-mechanical world. Things are recognized by the events they produce when they interact with a second object. An electromagnetic field is defined by its effects, for instance, on test charges, magnets, or photographic plates. It will produce events by accelerating charges or creating an image on a photographic plate. Events are at the basis of observable phenomena and states characterize the dispositions of quantum systems to exhibit certain behaviors. Both categories are closely interconnected. We prepare a system in a certain state by subjecting it to a complete set of compatible measurements. Suppose that we have a beam of electrons whose states we do not know. We can prepare the electrons with spin oriented in a certain direction $z$ and pointing upwards and in direction $z$ and pointing downwards by subjecting them to measurements with a Stern--Gerlach apparatus that separates the electrons in direction $z$ into two beams. One that deviates downwards that we call $|z,down>$ and another that deviates upwards $|z,up>$. If all the electrons that deflect downward are blocked with a photographic plate, we end up at the end of the day with a beam of electrons in the $|z,up>$ states. The process is based on a set of perfectly defined operations and the production of events in the lower beam. Electrons that are not intercepted by the measurement device are in a pure state $|z,up>$. Independently of the interpretation of the measurement process, this experimental setup based on ostensible observations provides us with systems in certain states.

 \subsection{Naturalism and epiphenomenalism}

Naturalism is a metaphysical view about the nature of reality and our knowledge about it \cite{decaro} . In its various versions, it maintains that reality is completely analyzable in scientific terms. In most cases, scientific naturalists assume, like Galileo, that reality is what the natural sciences say and nothing more. Scientific naturalism is the thesis that there is no other justification for our beliefs than those that the sciences can provide. Things that do not appear to admit a scientific description, such as qualia, normativity, responsibility, or freedom, must be described by science or explained away by science. An important form of scientific naturalism is physicalism. 

Naturalistic positions have repeatedly succumbed to the temptation of considering the empirically supported descriptions provided by physics to be complete. Physicalism asserts that everything is physical or is deduced from a physical basis. According to Papineau \cite{papineau}, the rise of physicalism is due to the causal closure of physics, which establishes that every physical effect has a physical cause. He himself recognizes that in the case of quantum mechanics causal closure must be understood differently and establishes that ``the probabilities of all physical events are completely determined by previous physical events." But he says that the distinction is basically irrelevant and ignores it in the rest of his argument because it would be an unnecessary complication to take it into account. This attitude towards the change in the notion of causality in quantum mechanics calls into question the type of physicalism adopted by Papineau. When a quantum event takes place, there is, in principle, a choice between the possible behaviors predicted by the theory that has no cause. The appearance of a dot on the photographic plate of the double slit experiment is not preceded by any physical cause that explains why it appeared there and not elsewhere on the plate \cite{persons}.

Determining the probabilities with which a certain event may occur is not enough to exclude in every physical situation another cause that could lead to the observed outcome. A widely held attitude among physicalists is that the determination of probabilities excludes any nonphysical effect. They do not consider the possibility that these effects could have an origin that transcends physics while respecting the probabilities assigned by quantum mechanics. Unless the latter possibility can be excluded, quantum mechanics makes the argument circular: causal closure in quantum mechanics would result from its determination of the probabilities of the results of a quantum measurement since the occurrence of a particular event could only be random in a physical world. Thus, the causal closure that would make physicalism plausible must, in the quantum case, be presupposed by assuming that the results are always random and therefore senseless \cite{persons}. It is important to note that randomness does not necessarily imply absence of meaning. It is possible to transmit information or meaning with a succession of signs that correspond to a certain probability distribution. For example, in English, letters appear with a certain probability. This does not prevent us from using the language to communicate.  A probabilistic theory like quantum mechanics does not exclude the existence of nonphysical causal properties hidden behind a series of events or measurement results.

Chalmers has analyzed the possibility of explaining the hard problem of consciousness in physicalist terms. To do this, he resorts to the notion of supervenience \cite{chalmers}. This is a notion based on the idea that one set of facts can completely determine another set of facts. Schematically, ``a set of properties T supervenes on another set B just in case two things cannot differ with respect to the properties T without also differing with respect to their properties B. In slogan form, ``there cannot be a difference in T without a difference in B'' \cite{superveniencesep}. The properties of T and B represent two levels of description, with B being the most fundamental level ---or bottom level--- from a physical point of view, and T being the higher level, or top level. A purely physicalist explanation would adopt a particular form of supervenience called reductive which requires that the T-properties can be logically deduced from the B-properties. Chalmers has given a series of arguments to show that it is not possible for the phenomenal to reductively supervene on the physical. In our opinion, the most powerful and sufficient argument to demonstrate that such a form of supervenience is not possible is the following, based on the structuralist approach to physics. The objects about which physics makes predictions are either mathematically describable objects or are complex processes like the life processes, whose functioning, although specifiable, is too complex to admit a useful mathematical description. But this kind of explanation of how something works simply leaves out what a conscious experience means. The phenomenal aspect does not result from a particular way of functioning, but from the qualitative aspects present in human experience. ``Although conscious states may play various causal roles, they are not defined by their causal roles. Rather, what makes them conscious is that they have a certain phenomenal feel, and this feel is not something that can be functionally defined away'' \cite{chalmers} p.105. Up to now, we have discussed a form of reductive physicalism that only admits the logical supervenience of higher structures and behaviors. This kind of physicalism does not seem to allow one to solve the hard problem.  

An alternative is to explore other forms of supervenience and relax the physicalist requirement, which, as we have seen, is not justified given the role that physical theories play as mathematical descriptions of a reality that transcends them. Implicitly accepting the causal closure of physics, Chalmers \cite{chalmers} has proposed a solution to the hard problem of consciousness based on a new form of supervenience that he calls natural supervenience because it is based on new laws that connect the physical with the phenomenal. It is therefore a form of property dualism. Within this approach, “Consciousness is a feature of the world over and above the physical features of the world... It remains plausible... that consciousness {\em arises} from a physical basis, even though it is not {\em entailed} by that basis." \cite{chalmers} p.125. It is important to note that this proposal abandons the idea that physics is sufficient to describe the totality of what exists, admitting that new laws and properties are necessary. However, it assumes the typical causal closure of classical physics. Therefore, the new {\em psychophysical} laws cannot interfere with physical processes and will only be rules of supervenience. This seems contradictory; if our phenomenal experiences cannot affect the causal description provided by physics and are therefore epiphenomenal, how can we explain why we feel driven to reflect on them or feel that they are essential in our valuation of life? An epiphenomenal vision would also not contribute to understanding normativity, responsibility, or freedom as we mentioned above. In the next section, we will introduce an alternative metaphysical hypothesis to the forms of naturalism presented here that seems more suitable to explain these problems and is fully compatible with physics.

\section{Regularism} 

Naturalism, both in its reductionist version, unable to explain the phenomenal aspects of consciousness, and in its version of natural supervenience, does not provide a satisfactory explanation of our conscious experiences. In the best case, it allows for an epiphenomenal description. We will show here that there is another alternative that, without questioning the value of science as a rigorous vision of natural behaviors based on evidence, is open to the totality of the capacities of the human and animal mind. 

The world accessible in third person presents regularities. The task of physics is the representation of these regularities in mathematical terms. The successes of this representation have led many to replace the concrete world that is revealed in our experiences with its mathematical idealization. 
We call regularism \cite{hospitable} the position shared by philosophers like Wittgenstein or van Fraassen and physicists like Hertz, Mach, or Hawking that distinguishes between the mathematical laws and the universe that they describe. Scientific representations have to do with mathematical structures and their correspondence to observable phenomena. At no point are the structures supposed to require any kind of similarity between the manifest and the scientific \cite{consciousness3}. Regularism is limited to establishing that the regularities that physics identifies in its description of processes and phenomena are fulfilled without exception, without asserting that nature is interchangeable with or exhausted by the physical description. There is, as we will see, a fundamental difference between a regularism based on classical physics and one based on quantum mechanics, since in classical physics there is causal closure, and in quantum mechanics not. Classically, current events are totally determined by the past, whereas in a quantum world, such as the one we inhabit, the determination is only probabilistic. As regularism does not assume that the world is exhausted by the physical, a probabilistic description is open to novel causal powers. 

An example could be the causal powers resulting from the phenomenal realm that could have a fundamental role in the natural world as assumed, for instance, by panpsychism. The regularism we have previously introduced establishes that regardless of how we account for the totality of observed phenomena including consciousness, the regularities ---the fundamental physical laws--- are always fulfilled without exception. Reducing consciousness to physics is unnecessary and probably impossible. But understanding the behavior of the natural world in a third-person perspective provides a starting point to understand the physical basis of consciousness; the brain processes that make consciousness possible. 

Given the fact that physics only describes processes in relational terms explaining how certain entities are related to others, it gives no information about the nature of what is related by these causal processes. As Wishom \cite{wishom} p.57 notices ``For those who accept structuralism about physical theories and hold that the natural world has an underlying nature that (at least at present) lies outside the scope of our knowledge, the question of this scientifically inscrutable nature emerges. ...After all, if there is indeed more to the natural world than can be fully and adequately described by physical theories, then we could simply chalk up our inability to find a place in the natural order of consciousness to our ignorance rather than to any deep divide in reality.''  Regularism leads us to think about the nature of the entities that are hidden behind the observed physical behaviors. The physical description tells us very little about the internal aspects of the entities to which it refers. It can tell us how these entities relate among themselves, but nothing about their intrinsic properties. Even properties like mass and charge are defined relationally. As we have privileged access to at least some of the intrinsic properties of our brain through our experience, it is natural to assume that they are at least partially phenomenal and widespread to any physical system. Therefore, we could consider these intrinsic properties as protophenomenal, leading, when suitably combined, to phenomenal properties like the ones we access through our experience. 

This would allow us to explain, for instance, why consciousness plays a central role in our life, transcending the previously discussed notion of natural supervenience. The latter,  as we have seen, is an epiphenomenal account of our inner experiences. If, on the contrary, the contemporary dominant physicalism were correct and the ultimate nature of the world was purely physico-mathematical, there would be no space for the exercise of responsible and creative freedom, and the indeterminacy allowed by quantum theories would only yield ``aimless random processes" \cite{kant}, see also \cite{mitchell}. It is worthwhile asking why the determinist hypothesis has persisted in force, although the fundamental theory of physics, quantum mechanics, is indeterministic. The fact is that, despite the success of quantum mechanics, worries about the possible determinism of human behavior have persisted. In the first place, there have been controversies about the interpretation of quantum mechanics and the existence of some minoritarian interpretations that are deterministic.  The resistance that some distinguished physicists like Einstein opposed to the quantum description is today totally superseded, and free-will theorems like the one from Conway and Kochen \cite{coko} show conclusively the random character of quantum theory. Nevertheless, the lack of consensus on how to interpret quantum mechanics has led many philosophers to continue to think in deterministic terms until the situation is finally resolved.  An additional support to the deterministic position has come from the belief that the quantum indeterminacies cannot be relevant in hot and humid media like the brain. It is precisely this point that makes the results questionable in light of new discoveries that prove that quantum mechanics is possible in biological media such as the brain, where several examples of the role of quantum effects have been found beyond molecular operation \cite{mcfadden}, \cite{ap} \cite{consciousness3}. Furthermore the kinematics introduced in the axioms of quantum mechanics and most of the realist interpretations of the theory coincide in the viewpoints required to adopt an event ontology, being particularly appropriate to account for conscious phenomena.

Another objection often raised to the possibility of any influence of the mind is based on the conservation of energy. For instance, John Searle p.42 \cite{searle} say : ``physics says that the amount of matter/energy in the universe is constant, but substance dualism seems to imply that there is another kind of energy, mental energy or spiritual energy, that is not fixed by physics. So, if substance dualism is true, then it seems that one of the most fundamental laws of physics, the law of conservation, must be false''.  Although we do not consider that the quantum regularism presented here is a form of dualism, it would seem that this objection could also be raised. This argument is based on several false assumptions about the energy conservation laws and the nature of the action of the mind that do not apply to quantum mechanics \cite{persons}. 

The probabilistic character of quantum mechanics and therefore its causal openness is manifested during the measurement processes that lead to the occurrence of events. From the point of view of quantum mechanics, conservation laws of energy or other classically conserved quantities usually are related with magnitudes called constants of the motion that are not conserved by measurements. Although energy is conserved by the Schr\"odinger evolution, energy is not conserved during measurements. The Schr\"odinger law describes the evolution of a subsystem in isolation, a measurement involves a system interacting with other systems. Energy may be exchanged between the measuring apparatus and the measured subsystem. What are conserved are expectation values of the constants of the motion. In other words, the probability rules of quantum mechanics ensure that if one computes the mean value of the energy of a system in a certain state by repeating the measurement many times, this mean value remains unchanged by the measurement. Here we assume that novelty arises from events occurring during measurements and is therefore consistent with the probabilistic assignment made by quantum mechanics \cite{Carroll}. Regularism is a position strictly compatible with scientific results where qualia, responsibility, or freedom could in principle take place without being reduced to illusory manifestations of a causally closed physical universe.

\section{Quantum Panprotopsychism}

Panpsychism is the thesis that some fundamental physical entities have mind-like elements. The idea dates back to antiquity and forms of panpsychism were also raised in modern times by Spinoza or Leibnitz, but it was taken up with particular interest at the beginning of the twentieth century by William James, Alfred North Whitehead, and Bertrand Russell. James adopted a form of panpsychism close to Spinoza's dual-aspect view and subjected it to a very sharp criticism, being one of the first to raise one of the most serious objections to this point of view: the combination problem. Whitehead proposed a radical reform of our conception of the fundamental nature of the world, placing events as its core feature \cite{seagersep}. His notions of actual occasions and prehensions are very well suited for the introduction of a notion of subject in an ontology of states and events as the one resulting from quantum mechanics, and we will come back to them. But the notion of panprotopsychism that we will adopt here is closer to Russellian monism, in which panprotophenomena play a central role and physics describes how they are mathematically structured.

Usually, in panprotopsychism, a kind of mind-like element is attributed to elementary particles such as quarks and photons. As quantum physics teaches us, this kind of attribution is incorrect; the fundamental concepts of quantum mechanics according to its axioms are systems in certain states that produce events in their interaction with other systems. Chalmers somehow recognizes this \cite{chalmers3}. He says p.1: ``Perhaps it would not suffice for just one photon to have mental states. The line here is blurry, but we can read the definition as requiring that all members of some fundamental physical types (all photons, for example) have mental states." It is important to note that he uses mental states here to speak about what we have called mind-like elements. 

Regularism and an ontology of events based on quantum mechanics could provide \cite{consciousness3} the following interpretation: events in the external world are subject to a physical description but not necessarily exhausted by it. This is so because at least some events in our brain could be directly accessible as perceptions. The main difference between both forms of events is the way we access them: first-person access for the mental and third-person access for the physical.  This is a form of panprotopsychism, which is the doctrine that fundamental physical entities, such as events and states, have intrinsic features that, although not mental in the usual sense, when appropriately arranged, give rise to consciousness in complex creatures like us \cite{wishom}. According to regularism, the third-person description of nature does not exclude nuomenal aspects given the causal openness of quantum mechanics.

One could say that the panprotopsychism proposed here is a kind of protophenomenal regularism. It is protophenomenal because fundamental physical entities would have intrinsic phenomenal aspects of a primordial type, which do not necessarily imply a subject with a first-person perspective, and it is regularism because the third-person aspects of fundamental entities are only restricted by their causal connection with other entities that obey physical laws. According to this view, some kind of mental life would not only be enjoyed by humans and some animals, but the primordial elements of the mind would be common to many physical entities. This position was advanced by Russell in 1921: “I think.. that an ultimate scientific account of what goes on in the world, if it were ascertainable, would resemble psychology rather than physics... such an account would not be content to speak, even formally, as though matter, which is a logical fiction, were the ultimate reality. I think that, if our scientific knowledge were adequate to the task, which it neither is nor is likely to become, it would... state the causal laws of the world in terms of…particulars, not in terms of matter. Causal laws so stated would, I believe, be applicable to psychology and physics equally; the science in which they were stated would succeed in achieving what metaphysics has vainly attempted, namely a unified account of what really happens, wholly true even if not the whole of truth, and free from all convenient fictions or unwarrantable assumptions of metaphysical entities” \cite{Russellmind} 2005 p.185. In his analysis of matter, Russell makes clear that he shares the view that physics only describes structures and causal connections and that matter in a given place are all the events that are there and goes on to say that such vision of matter implies that: ``We no longer have to contend with what used to be mysterious in the causal theory of perception: [how] a series of waves of light or sound... suddenly produce a mental event apparently totally different from them in its character \cite{Russellmatter} p 185". Consequently, the physical and the protopsychic would be nothing more than two ways of approaching the same object: the event. His notion of an event is closely related to its experience; for example, he says: ``Everything in the world is made up of events... An event, as I understand it... is something that occupies a small finite amount of space-time... When I speak of an ‘event’ I do not mean anything outside the way. Seeing a flash of lightning is an event; so is hearing a tire burst, or smelling a rotten egg, or feeling the coldness of a frog'' \cite{Russellmatter} p. 222, describing in this way events by their perceptions.

\subsection{The fundamental entities of the world: events and states}

Panpsychism is an attempt to assign phenomenal properties to the fundamental entities represented by physics to ``recover an ontological continuity in nature without forgetting its most obvious manifest feature [consciousness]” \cite{bitbol} p. 342. Attempts to base an ontology on events have a long history
that was reinforced by relativity and quantum theory. We consider that there is a minimal form of realism that should be accepted to make sense of our physical theories. It is the one limited to assign reality to those entities directly present in the correspondence rules of a theory \cite{Jammer}, which allow us to connect the formalism with our experience to verify or falsify the theories. The axioms of quantum mechanics refer to primitive concepts such as systems, states, events, and the properties that characterize them. The use of these concepts suggests that the theory should admit an ontology of objects (understood as systems in given states) and events. There are entities that even defenders of empirical structuralism such as van Fraasen admit as real, they are the correlates of observable phenomena.  Based on this ontology, objects and events can be considered the building blocks of reality. Objects will be represented in the quantum formalism by systems in certain states. In an event interpretation like the ones we are considering, events are the actual entities, while states represent dispositions to produce events. In quantum mechanics, a state is mathematically represented by a vector in a Hilbert space and characterizes the disposition of a system to produce events in other systems.  The basic idea of a measurement is the occurrence of a macroscopic phenomenon, that is, something capable of reaching perception \cite{consciousness3}. Thus, as noticed by Omnes, the measurement of a property of a microscopic object implies generating a phenomenon, in other terms producing an event. The process of detection of photons by dissociation of silver bromide in a photographic plate leading to a cascade effect that produces the accumulation of millions of atoms of silver is an example of the production of an event: The appearance of a dot on the photographic plate \cite{eventontology}. It is an example of a macroscopic event that contributes to making the world accessible to our senses \cite{Omnes}. As Whitehead \cite{whitehead} recognized: ``the event is the ultimate unit of natural occurrence.'' 

The event ontology has the attractive feature of eliminating the divide between the mental and material world. As Russell \cite{Russellmatter} p.275 pointed out, ``if we can construct a theory for the physical world which makes its events continuous to perception, we have improved the metaphysical status of physics.” According to his view, we need ``an interpretation of physics that gives due place to perceptions” \cite{Russellmatter}.  As we put it in \cite{hospitable}: An ontology of events based on quantum mechanics could provide this interpretation. Events in the external world are subject to a physical description but not necessarily exhausted by it, since in quantum mechanics there is no causal closure, while at least some events in our brain could be directly accessible as perceptions. The main difference between both forms of events is the way we access them: first-person access for the mental and third-person access for the physical. 

Although the position presented here shares strong similarities with the one proposed by Russell known as neutral monism, it differs in several aspects that we will discuss next. He seems to lean toward an alternative ---that we share--- that sustains that the basic properties of the world that are neither physical nor phenomenal, but give rise to both. ``From their intrinsic natures in combination, the phenomenal is constructed; and from their extrinsic relations, the physical is constructed" \cite{chalmers} p. 155.  However, the quantum ontology of objects and events that we propose here differs from the notion proposed by Russell in that it includes, in addition to events, systems in certain states \cite{consciousness3}. If the quantum events have a phenomenal intrinsic aspect, does something similar occur with the systems in certain states that we have called objects?  What we describe as passions, emotions, and the resulting voluntary behavior could be related with what we have described up to now as dispositional quantum states. Precisely this volitive aspect is what could be associated with states that, in quantum mechanics, have been related with dispositions or tendencies to produce events. Notice the similarities that exist between the characteristics of mental phenomena, such as emotions, intentions or feelings, dispositional and private, and those of the states. The latter are dispositional and inaccessible by isolated measurements. As the measurement changes the states, they are inaccessible and, therefore, private and may have an internal aspect like the one we assume here. 

To characterize the disposition of a state to act on any other system is to give its most complete mathematical description given by a vector or a density operator in a Hilbert space \cite{consciousness3}. States are private in the sense that one cannot precisely determine in which state a system is by measuring it. One needs an ensemble of identical states and measurements on each member of the ensemble to have a complete determination of the states. Due to the no-cloning theorem, it does not make sense to have an ensemble of identical mental states; they are physically inaccessible to external observers. It is important to note that states would simultaneously have psychological and phenomenal aspects. Indeed, it is the state that dictates the tendencies or dispositions that physics studies that lead to the production of events. Quantum mental states fulfill a double function: Insofar as dispositions determine probabilities for certain actions, which are studied by psychology. Insofar as they possess an internal aspect expressed in inclinations or emotions, they are phenomenal. When one considers non-local systems like particles in entangled states whose components occupy different positions in space-time, it is not possible to speak of a state at a given time, since that is a notion that depends on the Lorentz reference frame chosen. However, if the state is defined by its disposition to produce events, one can rigorously show \cite{gapo} that such a disposition is uniquely defined and that the state in the Heisenberg picture only changes when events take place. The disposition to produce events, even if they are spatially separated in the sense of relativity ----that is, not causally connected--- is an objective property of states. This underscores the objectivity of states when characterized in terms of dispositions. They have the same ontological status as events that are the ostensible aspect of reality.

A system in a given quantum state can only be fully prepared by measuring a complete set of compatible observables: depending on the system, the number of observables required may change. As we shall see, in simple cases, a measurement is enough to determine the state.  A state corresponds to each result of the complete set of measurements; in other words, it is correlated with a set of event observations. Both phenomena and specific states can be considered correlates of conscious contents. However, most states are not completely accessible to our knowledge because many measurements are needed to determine the state of a physical system and measurements alter the state. Note that objects are only observable through their manifestations in events resulting from their interaction with other objects \cite{consciousness3}. This does not mean that objects are mere hypothetical entities. We can operate on an object, for example, a hydrogen atom, and place it in a state that interests us, such as placing it in an excited state whose energy we know. But these operations are always carried out based on information provided by events: the only entities accessible in the third person to our consciousness. Quantum field theory allows us to demonstrate that the electromagnetic field, like others, also admits an ontology of objects and events, and existing versions of quantum gravity that unify general relativity with quantum mechanics show that the same ontology is applicable to spacetime \cite{brha}.

We want to concentrate here on investigating to what extent we can advance a panpsychist description that incorporates the paradigm shifts that a quantum vision implies. The problem is not merely philosophical, but has an empirical dimension, requiring the existence of quantum behaviors that are difficult to maintain in the brain and that we need to identify \cite{consciousness3}. We are interested in analyzing to what extent consciousness can be understood from a third-person perspective, such as that provided by science. Whether it is possible for a physical approach to brain functioning to be at least compatible with the existence of a phenomenal world, a physical description of the brain must at least explain how third-person knowledge of the brain is compatible with our conscious experience.

\subsection{Quantum Emergentist Panpsychism}

The usual philosophical treatment of emergence is extremely unsatisfactory and is based on a world view based on classical physics that has been largely outdated. David Chalmers \cite{chalmers3} p. 8, identifies two forms of panpsychism, constitutive and non-constitutive or emergent. ``Constitutive panprotopsychism is roughly the thesis that macroexperience is grounded in the protophenomenal properties of microphysical entities. That is, all phenomenal truths are grounded in protophenomenal truths concerning these entities.'' Therefore, it is a classic panprotopsychism in physical terms. In classical theories, like Newtonian mechanics or general relativity, the physical counterpart of the states of a system is given by a particular set of data correlated with events. For Newtonian mechanics their empirical counterparts are the positions and velocities of the component particles of the system. The properties of each particle of the system at a given instant. Any observable quantity is a function of these properties. If one adopts an ontology based on classical physics, one must adopt a constitutive panprotopsychism: every property of the whole is a function of the properties of each of its parts.
As Seager\cite{seagersep} observes, if one adopts this point of view, ``the fact that I am conscious consists primarily in the fact that certain particles in my brain are arranged or interacting in a certain way... This is the form of panpsychism that suffers most acutely from the combination problem.''

Non-constitutive panpsychism holds that there are micro- and macro-experiences, but the micro-experiences do not ground the macro-experiences. This form of panpsychism is impossible if an ontology based on classical physics is adopted, but, as has been observed \cite{lulegues, healey, teller}, it is the natural ontology that results from quantum physics. 
Quantum mechanically, there exists what Healey \cite{healey} Sec.4 calls ``Physical Property Holism". It establishes that there are physical objects ``not all whose qualitative intrinsic physical properties and relations supervene on qualitative intrinsic physical properties and relations in the supervenience basis of their ``basic physical parts."” The emergence of new properties of the whole in a quantum world, where events and properties play a fundamental role, is a crucial manifestation of the ontological novelty and non-separability most characteristic of quantum phenomena in entangled states \cite{consciousness3}. These behaviors of systems in entangled states allow wholes to have properties that cannot be explained in terms of the properties of their parts. In a model of two entangled spinning particles in a singlet or triplet state, the components along certain directions of the total spin of the two-particle system are an example of a property that does not supervene from its parts. In fact, the parts do not have any defined component of the spin. The novelty of these properties can play an essential role in the emergence of consciousness in composite systems with many components. The number of emergent properties that the total system may have that do not supervene from the properties of their parts grows exponentially with the number of parts. For example, a system of $n$ entangled spinning particles may be in $2^n$ different pure states with different properties of the total system and different dispositions or causal powers to produce events in other systems.

Paul Teller \cite{teller} p.76 has given a different characterization of the behavior of entangled states. He defines as local supervenience the following feature of any classical system: ``the world is composed of diverse physical individuals that possess non-relational (i.e., monadic) properties; and... all relations existing among such entities supervene upon the non-relational properties of the relata.” Quantum systems exhibit ``relational holism”, that is, the standpoint that admits that there exist relations that do not supervene from the non-relational properties of their relata \cite{consciousness3}. Entangled states in quantum mechanics are of this kind; that is, they exhibit relational holism. In the example of entangled spins, there are states with no non-relational properties associated with the spin components of their component particles, but the total system may have a number of different sets of properties that grows exponentially with the number of particles. Note that this type of emergence is a direct consequence of the physics of entangled many-body systems. In what concerns the panpsychist hypothesis, this does not imply that human/animal consciousness is not caused by microlevel consciousness but that the fundamental objects, states, and events of entangled composite systems would present new properties and causal behaviors that do not result from adding those of their parts \cite{lulegues,consciousness3}. 

Karakostas \cite{karakostas} p. 268, compares the resulting quantum picture with the Humean ontology as follows: ``From the perspective of Lewis’ ---Humean--- metaphysics, if one can determine the intrinsic qualities of particular events or atomic objects in space and time, then one can describe the world completely. Quantum mechanics, however, is not in conformity with Lewis' atomistic metaphysical picture, which depicts a world of self-contained, unconnected particulars that exist independently of each other... the consideration of physical reality cannot be comprehended as the sum of its parts in conjunction with the spatiotemporal relations among the parts, since the quantum whole provides the framework for the existence of the parts... their entangled relation does not supervene upon any intrinsic or relational properties of the parts taken separately. In fact, this is the characteristic that makes quantum theory go beyond any mechanistic or atomistic thinking.” {\em This behavior is a consequence of the kinematics of the quantum theory and completely independent of the solution that one would like to adopt for the measurement problem.}

The ``holistic" behavior of many quantum systems that are non-separable is not something exceptional. It is the most common occurrence in composite systems whose components interact or have interacted in the past. For example, the electrons in a multielectron atom are entangled. It is a generic property: interactions typically yield entangled states for multi-component systems. Quantum entanglement and non-separability are closely related with coherence. Much of the world accessible to our daily experiences is separable precisely because its quantum nature is masked by the pervasive phenomenon of environmental decoherence that generates the observed classical behaviors.

Non-separability and holism become more important as the more particles the systems involve. Only a subset of that exponential number of states that grows linearly with the number of components presents classical behaviors. It is the number of entangled states that grows exponentially. It is related to the number of independent coefficients necessary to determine the entangled state of the $n$ particles. This growth in the possible behaviors of the quantum systems and the possible entangled states, is at the basis of the appearance of novel properties and top-down behaviors of emergent systems. Their philosophical implications can be recognized when the appropriate ontology is put into action. Most of the events, described by the projectors in the Hilbert space, correspond to entangled states and are differentiated from the simple sets of events of the nonentangled parts in that they present new emergent properties. They therefore have the possibility of adopting a very large number of properties and states.  The assumption that it is not necessary to analyze the mind-body problem from a quantum ontology is not well-founded. First, because there are quantum models of cognition compatible with the physical conditions of the brain, and second, because it cannot be ignored that quantum mechanics provides an ontology that is much more rich and adjusted to phenomenal evidence.\cite{consciousness3}.

This type of emergence is sufficient to resolve most objections related to the combination problem \cite{chalmers2}, but it requires that entangled quantum structures coupled with the nervous system be identified in the brain \cite{consciousness3}. It is important to note that quantum emergence does not require radical forms of emergence, where the 'emergent' properties of a complex system and its corresponding states cannot be intelligibly derived from the properties of its parts. Here intelligible means mathematically derivable by the same physical theory that describes the states and events of its parts. They simply result from the fundamental laws of interaction when parts are put together. The states of entangled quantum systems have top-down causal properties that cannot be deduced from the states of their parts, and as we have emphasized here and in previous works \cite{lulegues,consciousness3} entangled systems possess properties even when their parts do not possess any defined properties. 

This kind of emergence could not by itself explain the phenomenal properties of consciousness; it is only in the panpsychist or panprotopsychist context that it leads from elementary phenomenal forms to the higher ones found in human beings and other mammals.
The kind of panprotopsychism we adhere to here, based on regularism, is a form of Russellian monism \cite{chalmers3}. This view takes its name from Russell’s insight, in The Analysis of Matter and other works, that physics reveals the relational structure of matter but not its intrinsic nature. This is in line with the kind of structuralist interpretation of physics adopted here following van Fraasen. Russellian panpsychism is the view that some quiddities are microphenomenal properties. As Chalmers puts it \cite{chalmers3} p. 9, ``According to this view, classical physics tells us a lot about what mass is ---it resists acceleration, attracts other masses, etc.--- but it tells us nothing about what mass intrinsically is. We might say that physics tells us what the mass role is, but it does not tell us what property plays this role." The physical description is, therefore, compatible with the intrinsic phenomenal aspects. 

Quantum indeterminacy manifests itself only in measurements. The events that occur in measuring devices can only be predicted probabilistically. If there is any effect of the phenomenal on the physical, it would be through choices of events that occur in the measurement processes. Therefore, the possible physical impact of the phenomenal inner experiential aspect on our brain and indirectly on our actions could only occur in these processes. In a panprotopsychist context, they would correspond to choices of the state of the entangled system to which we attribute the experiences coupled with the measuring device, which in this case would be the neural system. This permanent manifestation of the phenomenal in the physical world is the basis for the constitution of our perceptual experiences, which, as Merleau Ponty observed, necessarily involve a learning process in which our body plays a central role in the coordination of the signals of our sense organs \cite{merleau}. It should be noted that this process does not imply any form of dualism since there would exist a single form of reality with a phenomenal aspect whose physical behavior is ordered and limited according to quantum principles.

\section{The combination Problem of panprotopsychism}

Any attempt to understand panpsychism by accepting the structural description of the world resulting from classical physics is problematic. It fails to explain the emergence of complex minds from the right kinds of interactions among multitudes of entities exhibiting micro-consciousness. A problem without apparent solution in this context is to explain how any combination of more basic mental states or qualities similar to the mind could somehow add up to conscious mental lives like ours \cite{wishom}. The problem was first raised by William James \cite{james} p. 160,
``Where the elemental units [which compose complex minds] are supposed to be feelings, the case is in no way altered. Take a hundred of them, shuffle them and pack them as close together as you can (whatever that may mean); still each remains the same feeling it always was, shut in its own skin, windowless, ignorant of what the other feelings are and mean. There would be a hundred-and-first feeling there, if, when a group or series of such feelings were set up, a consciousness belonging to the group as such should emerge... but they would have no substantial identity with it, nor it with them, and one could never deduce the one from the others, or (in any intelligible sense) say that they evolved it.... Private minds do not agglomerate into a higher compound mind.” 

In other words, James thinks that, as in classical physics, the properties of the parts remain unchanged in the whole and that the properties of the whole are nothing more than functions of the properties of its parts ---they are given by them---.  From this assumption, we observe that it is no more intelligible to us how any number of simple subjects of experience could combine to produce a new complex subject of experience than it is for a combination of insentient materials to do so \cite{wishom}. The objection is therefore, as Seager \cite{seagersep} points out, that it “is very difficult to make sense of: 'little' conscious subjects of experience with their micro-experiences coming together to form a 'big' conscious subject with its own experiences.'' This problem that applies to every aspect of our conscious life, both in relation to our subjectivity and our perceptions, is considered by panpsychists as one of the most pressing challenges to their view. Due to this problem ``In the eyes of most contemporary philosophers of mind, panpsychism remains a radical and extravagant answer to the problem of integration'' \cite{wishom} p. 61.

Chalmers\cite{chalmers4} has analyzed the forms that the combination problem adopts by identifying three subproblems based on different aspects of phenomenal consciousness: its subjective character ---it always involves a subject---, its qualitative character ---it involves a great number of different qualities---, and its structural character ---it organizes in very complex structures---. They are respectively called the subject combination problem, the quality combination problem, and the structure combination problem. 

The subject combination problem is about how microscopic subjects of experience combine to yield macroscopic ones such as ourselves. One can identify here different subproblems that we would like to address: what are subjects of experience, how they combine, and how the subject summing problem can be solved. The subject summing problem consists of understanding how microsubjects may contribute to the appearance of a macrosubject. Chalmers presents the problem in a different form ``given any group of subjects and any other subject, it seems possible in principle for the first group of subjects to exist without the further subject. If so, then no group of microsubjects requires the existence of a macrosubject” \cite{chalmers4} p. 5. Once again, the problem stated in this form is posed by assuming a form of classical supervenience where the properties of the whole can result only from those of its parts that preserve their subjective aspects.

The quality combination problem is roughly the following: How do the basic microscopic properties whose intrinsic nature is described by physics and are relatively few combine to lead to the incredible variety of qualities that result from our different senses? A particular form of combination problem is {\em the palette problem}: how do the wildly different qualities of sensory organs --odors, colors, or sounds--- result from a few basic physical properties? Seager \cite{seagersep} summarizes metaphorically the palette problem as follows: ``How is it that the richly painted canvas of human experiences is produced from such a small palette of paints?''

Human conscious experience is rich in not only qualities but also structure. Our experiences seem to present a rich structure including spatially organized optical and auditory fields and including kinesthetic perceptions in an integrated whole \cite{chalmers}. The structural mismatch problem has been summarized by Chalmers as follows (p.5) ``Macrophysical structure (in the brain, say) seems entirely different from the macrophenomenal structure we experience. Microexperiences presumably have structure closely corresponding to microphysical structure…, and we might expect a combination of them to yield something akin to macrophysical structure. How do these combine to produce a macrophenomenal structure instead?'' 

In a previous paper \cite{consciousness3}, we have analyzed the quality combination problem and advanced some considerations about the structural combination problem. Here we will review the quality combination problem and discuss the two remaining problems concerning the structures and subjects. As we shall see, the solution to these problems sheds light on the types of brain functioning that should be expected and need to be identified.  

In that paper, we have also recently observed that if a quantum ontology is adopted, the combination problem is diluted, at least in relation to some aspects of it. In the next subsection, we will review that solution to the quality and grain ---a subproblem of the structure--- combination problems to analyze in what follows how to resolve the aspects that are still pending in the structural problem and \cite{seagersep} the subject-summing problem.

\subsection{The Quality Combination Problem of Panprotopsychism}

We have already analyzed this problem in the previously mentioned paper; we briefly review the main ideas about the solution of this issue.  Recall that the quality combination problem consists in the understanding of how do microqualities combine to yield macroqualities? Here macroqualities are specific phenomenal qualities such as phenomenal redness (what it is like to see red), phenomenal greenness, etc. \cite{chalmers4}. Again, in the panprotopsychist version, heterogeneous qualities are being compared. If the world behaved classically, the problem would be how qualities that accompany certain physical events, such as a neuron firing, yield the qualities that appear at the end of processes that start in our sensory organs and are organized by our brain into complex perceptions. However, the preceding quantum analysis of states and events allows the problem to be posed in much more promising terms if there is a quantum system of entangled molecules or photons coupled to the neural system. 

The experienced qualities should therefore be correlated with the properties of physical macro-events and their corresponding abrupt changes of state. The quantum entanglement of states and their properties does not allow us to think that the properties or states of the whole are composed of properties or states of the parts as in classical physics. Instead, in a quantum system, parts lose their individuality, and \cite{lulegues,consciousness3} only the properties of the whole are defined, as observed by Healey and others. In other words, in a quantum system capable of maintaining its entanglement, macro-qualities would appear associated to properties of macro-events without leaving a trace of the properties or micro-qualities of the entities involved in its constitution of the system in an entangled state. The number of possible properties of macro events grows exponentially with the number of entities involved in the macrosystem.

The quality combination problem results from the incorrect prejudice in a quantum system of supervenience of the state of the total system on the states of the microsystems. As we saw when analyzing top-down quantum causality, and we have explicitly proved elsewhere \cite{lulegues,consciousness3} the union of individual entities in definite states does not represent the entire entity because of the existence of information in entangled systems about correlations that is not contained in the states and properties of the parts. The above-mentioned homogeneous experience results from a process involving large portions of the brain interacting with an entangled system of many particles. To enable a solution, like the suggested one with top-down causation, it is necessary to have a quantum system composed of many microscopic quantum components, such as molecules or photons, in entangled states that extend across macroscopic regions of the brain coupled with the brain's neural system. The component particles, when combined in an entangled state, lose their properties, and only the full system in this state may have well-defined properties with their corresponding qualities.

The most tangible manifestation of the prejudice of assuming that the properties of the whole come from its parts, a true hypothesis in classical physics but false in quantum physics, is posed as another subproblem linked to the quality combination problem that Chalmers \cite{ chalmers2} calls the palette problem. If the classical ontology were applicable, we could expect only a handful of microqualities, corresponding to a handful of fundamental microphysical properties. How can this limited palette of microqualities combine to yield the vast array of macroqualities? Here, clearly appear the prejudices resulting from classical physics that generate this problem. Indeed, again it is explicitly assumed in this case, corresponding to the constitutive panpsychism position, that the properties of the whole come from those of its parts and therefore grow linearly with their number. Quantum mechanics provides an exponential number of new states and properties in which those of the parts are not defined any more and gives a satisfactory and intelligible answer to the palette problem. The states and events in our brain can have the complexity of our emotions and perceptions given that they correspond to an exponentially high number of possible physical events. Returning to Seager's metaphor, quantum mechanics provides us with an exponentially large palette of paints to produce richly painted canvases of human experiences.

\subsection{The Structure Combination Problem of panprotopsychism}

The human conscious experience is not only rich in qualities, but also in structure \cite{seagersep}. The richness of our perceptions and their organization into complex spatial representations that combine data from different senses in our perceptions seem very different from the structure of the brain. This is the problem of how complex arrangements of vast numbers of discrete, spatially discontinuous protophenomenal but insentient materials could yield a homogeneous experience. Like the experience of a landscape, or perceptions that are unified, multifaceted, and structured both egocentrically and in terms of objects represented as being in the environment. The rich structure of consciousness results from, or at least coexists with, the apparently very different structure of the brain. This fact should be addressed by any form of panpsychism. A particularly pressing aspect of this problem called the `grain problem' \cite{maxwell, lockwood} is the concern that experiences appear fluid and continuous in a way that is at odds with the discrete and particularized structure of the brain's properties. As put by Chalmers \cite{chalmers4} p.5  ``Microexperiences presumably have structure closely corresponding to microphysical structure…, and we might expect a combination of them to yield something akin to macrophysical structure. How do these combine to yield macrophenomenal structure instead?'' The grain problem results from that incorrect prejudice in a quantum system of the supervenience of the state of the total system on the states of the microsystems, in particular, with the neural organization \cite{consciousness3}. As we saw when analyzing top-down quantum causality, the union of individual entities does not represent the entire entity. The above-mentioned homogeneous experience should result from a process of elaboration involving large portions of the brain interacting with an entangled quantum system of many particles. To enable a solution, like the suggested one with top-down causation, it is necessary to have a quantum system composed of many microscopic quantum components, such as molecules, in entangled states that extend across macroscopic regions of the brain and are capable of coupling with the brain's neural system. It will be from this quantum system that our phenomenal life will emerge.

On the other hand, there is a second aspect of the structure problem: “Our macroexperience has a rich structure, involving the complex spatial structure of visual and auditory fields, a division into many different modalities, and so on. How can the structure in microexperience and microstructure yield this rich structure?” \cite{chalmers4} p.5. This problem is intimately related with the problem of the structuring of the world in perceptions on which Merleau Ponty \cite{merleau} has reflected so much. This problem involves understanding the genesis of this capacity, essential for survival in the animal world. It seems clear that it must have resulted from a process of progressive biological adaptations guided by evolution. These adaptations cannot only be understood in physical terms since it is necessary to explain how the phenomenal conscious experiences are structured in these perceptions. In fact, the prejudice shared by most empiricists that sensations come first and perceptions are judgements about an already present sensation makes the structure problem unsolvable. This is because it assumes that this structure is already present from the very beginning. The origin of perceptions can be subject to analysis in evolutionary terms given their central role in the behavior and development of animals. What must be analyzed is not, in a panpsychist context, whether phenomenal consciousness can arise evolutionarily, but rather whether it is possible that more refined forms of perception and organization of sensory data imply adaptive differences that allow natural selection to operate. If this is the case, the ability to structure our conscious experiences organized in a life-world, in Husserlian sense, could evolve. Only regularism would allow us to understand how such phenomenal representations could be developed as more refined perceptions are required. In fact, interactions between the quantum system with a phenomenal counterpart and the nervous system would be necessary, which can occur due to a selection of events in the interaction of the state with the neural system that functions as a measuring device. Perceptions could thus evolve enriching  the  structuring of phenomenal representations \cite{Velmans}. Note that this scenario implies that, as our phenomenal experience indicates, our noematic contents would become richer and with them our capacity for adaptation greater.

In line with panpsychism, conscious intentionality reached new relevance with the development of superior animals, mainly birds and mammals. Motility, perception and emotion characterize the main innovations of animals. Their openness to the world is their central feature. To the urgency common to all living beings related with their basic concern to preserve and enhance their lives, animal life adds motility and ``the interposition of distance between urgency and attainment” \cite{jonas} p.101. The identification of prey requires the reinforcement of its abilities of perception, locomotion and keeping the prey under its attention and desire to achieve a non-immediate goal.

In the last decades a greater understandding of the evolution of cognition and consciousness has been achieved\cite{Butler}. Edelman et al. \cite{edelmantononi} have recognized that the most basic level of qualia, ``the collection of... subjective experiences, feelings, and sensations that accompany consciousness. For example, the ‘redness’ of a red object” \cite{Butler}, p. 443, are present in superior animals and a large amount of evidence has been accumulated regarding their perceptual abilities. In fact in the last few years the conviction has increased that consciousness is a widespread phenomenon in animal life. This conviction results fundamentally from the postulate that the correlation of complex cognitive abilities with higher-level consciousness present in humans holds for other species of birds and mammals and from their capacity to exhibit intentional behaviors  \cite{low}. ``Among the recently revealed high-level cognitive abilities of various species of birds are a number of manifestations of working memory, including delayed-match-to-sample, episodic memory, transitive inference, and multistability. Additional avian abilities include category formation, language use and numerical comprehension, tool manufacture and cultural transmission of tool design, theory of mind, and a high level of Piagetian object permanence.” \cite{Butler} p.445. 
Different phenomena related to perceptions have been also observed in animals, for instance the phenomenon of multistability, where ambiguous figures, such as the Necker cube, can be perceived as in one of two (or more) configurations, but no more than one of the configurations is perceived at a given moment, proving that the same intentional perceptual functioning has certain universality beyond the human beings \cite{Butler}.   Without a first-person perspective of what they observe, one could conjecture that the different stages noted in human perception by Husserl\cite{woodruff}, including the constitution of the noema, are also present in mammals and birds 
Category formation, which requires a high-level of cognitive ability, has been demonstrated in pigeons by Watanabe et al. \cite{watanabe}. Pigeons were first trained to discriminate pairs of abstract paintings by Picasso and Monet. In one case the Picasso works are reinforced and in the other set the Monet ones. They were then able to discriminate not only novel pairs of Picasso and Monet paintings but to react similarly to the Picasso paintings as to those of Braque and Matisse and to connect the Monet paintings to those of Renoir. The same type of clustering abilities have been observed in other birds and mammals \cite{Butler} p. 446. Butler reports  in that review other abilities as number recognition and human language recognition, also proving the ability of structure organization and pattern recognition. In what concerns the structuring of perceptions there is evidence of strong similarities between animal and human perceptual functioning. Particularly in the structuring of the information received by the different sensory organs into perceptions that organize and coordinate them spatially, as the recognition of patterns, natural and cultural objects, and many other phenomenal behaviors. Many forms of structuring of perceptions are already present in mammals and birds. We need to go further back in the evolutionary chain to understand how the capacity of spatial structuring evolves thanks to interactions between the phenomenal and the physical. Having solved the grain problem in quantum mechanical terms, which can be considered a prerequisite, the  problem of how to structure perceptions through evolution could be within the reach of cognitive sciences.
In summary, the structure combination problem presents two aspects that are, at least in part, scientifically accessible: the grain combination problem associated with the existence of entangled quantum states in the brain that must be identified to which their consciousness would be related, and a process of progressive development of biological perceptive capacities that would result from a physical-phenomenal interaction producing adaptive differences

\subsection{The Subject combination problem}

The subject combination problem is probably the oldest and most influential objection to panpsychism. It was formulated by William James in The Principles of Psychology \cite{james} as we cited above. 
He argues that experiences do not aggregate into further experiences, and the same happens with minds \cite{chalmers4}. If one has identified what protosubjects are and how they combine, one faces the subject-summing combination problem. It is difficult to see how the distinct conscious subjects combine into a single conscious mind. As in \cite{seagersep}, it is taken to be a challenge that the panpsychist must address in the long run, either by eventually coming up with an adequate theoretical account of mental combination, or at least by explaining why such an account is beyond our grasp. Again, we will claim that the problem arises from the incorrect use of a classical ontology. Sam Coleman \cite{coleman}, for example, has argued that subject summing is incoherent on the grounds that each subject has a viewpoint that excludes the viewpoints of all other subjects. The essence of our current point of view as a conscious subject is a matter not just of the conscious experiences that we are positively having but of the fact that we are having those experiences and not others.  He implicitly assumes that a combination preserves the individuals that make up the total subject, and therefore it would have to have both its experiences to the exclusion of all other experiences and the experiences of its components to the exclusion of all other experiences \cite{goff}. Objectors to solutions of the subject summing problem assume that the constitutive micropsychism holds and the facts about microsubjects wholly account for the existence of macro-level conscious subjects. Since, as we have noticed, in quantum mechanics the states and properties of the parts do not preserve their individuality, we claim that this view should be incorrect. However, we need a better understanding of what is the third person physical description that corresponds to subjects, especially to proto-subjects in elementary processes, and how they combine giving rise to new subjects.

\subsection{Subjects of experience and their quantum mechanical counterpart}

The first aspect that needs to be elucidated is the character of subjectivity and how to recognize it from its physical counterpart. Our starting point is inspired by Whitehead's analysis of subjectivity because it is well adapted to elementary physical processes involving the relationship between proto-subjects and objects. However, the application of his analysis to the ontology of quantum objects and events requires important modifications. A particularly relevant difference is in the notion of an 'object' in Whitehead's terminology, which differs significantly from the notion of a quantum object defined above. We use quotation marks to identify Whitehead's 'objects' that were later called eternal objects: chairs, electrons. They are not considered as dispositions at a certain time to manifest through events but as an entity independent of their position and time. In this regard, he says: ``Objects... are recognized as self-identical amid different circumstances; that is to say, the same object is recognized as related to diverse events” \cite{whiteheadknowledge} p.62. In fact, in his terminology, objects are the ancestors of eternal objects and events are the ancestors of actual occasions of experience \cite{christian}.

Let us schematically describe the basic elements of his analysis of subjectivity as presented in Adventures of ideas \cite{whiteheadaoi}. Whitehead begins by objecting to the usual distinction between subject and ‘object’ based on the opposition between knower and known. It does so based on the observation that a panpsychist vision such as the one implicitly adopted cannot be based on concepts that are only applicable to higher forms of consciousness such as the human one. Therefore, he chooses to assume that the most basic form of experience is emotional. More precisely, he establishes that \cite{whiteheadaoi} p.176: ``the basic fact is the rise of an affective tone originating from things whose relevance is given... about the status of the provoker in the provoked occasion."  More formally, he establishes p.176 that an occasion of experience is a 'subject' with respect to its special activity concerning an object, and anything is an object with respect to its provocation of some special activity within a subject. Such a mode of activity is called a 'prehension' ''. The word was introduced to describe an apprehension that may or may not be cognitive \cite{whitehead}. Three factors are involved: an occasion of experience with its corresponding prehension, a datum that provokes the origination of the prehension that is the prehended ‘object’, and there is the subjective form which is the affective tone of that prehension. The subject is the realm where the acts of experience occur; A is an act of experience means that A is a prehending entity. ``It does not imply that A is conscious of what it prehends or conscious of itself. Consciousness is a feature of some prehensions, but not all ” \cite{christian} p.19. Whitehead is, therefore, a panprotopsychist, he does not attribute consciousness to all entities in which prehensions or events occur, so it would be more appropriate to speak of proto subjects when we refer to the process of occurrence of events in physical systems. 

As we shall see in some detail in what follows, prehensions are closely related with quantum systems in certain states.  As noticed by Whitehead
\cite{WhiteheadProcess} p.19: ''A prehension... is referent to an external world, and in this sense will be said to have a 'vector character'; it involves emotion, and purpose, and valuation, and causation'' in the same way that states have dispositions to produce events.  Every time an event occurs, it is preceded by the prehension of a proto-subject. We know of at least one situation where quantum events occur: It is during measurements. Several realist interpretations of the quantum mechanical axioms clearly admit a treatment in terms of prehensions leading to events: the Copenhagen interpretation \cite{heisenberg}, the modal interpretation \cite{modal}, the Montevideo interpretation \cite{review}, Zurek’s existential interpretation\cite{zurek} among others.

There are two quantum physical processes where events play a central role, both physically and phenomenally, that we will analyze in detail because they are proto manifestations of occasions of experience related to two clearly distinguishable conscious phenomena. The preparation of a state where the occurrence of certain events places the system in a new state that evolves from that moment on; and the measurement of a state where an event results from a random choice that ends up changing or annihilating the previous state.  Schematically, in a preparation, the events put the subject in a certain state; in a measurement, the subject in a certain state produces an event in the object at the end of a process. In the former, a state is born with the event that characterizes its properties ---in particular its phenomenal qualities--- in the second it dies with the production of an event through its interaction with other physical systems. The first can be associated with a sensitive process, it produces a phenomenal quality, a proto perception; in the second, the phenomenal process culminates with an action, a proto volitional act.

Let us start with the analysis of subjects that end up producing events, and therefore measurement outcomes. Quantum states manifest themselves in the production of events during measurements. In this process, quantum indeterminism manifests itself. Every time an event occurs, it is preceded by the prehension of a proto-subject of experience. An essential element of this process, its climax, is the choice of one of the possible outcomes during the measurement process. Among other issues, it is fundamental to determine which physical entity is the third-person counterpart of the protosubject of experience involved in this choice of the event produced in the measurement device. In the measurement of elementary physical properties, such as the position of an electron, these choices are merely random. Let us analyze one of these measurement processes in detail to identify its different phenomenal elements which would result from the quantum ontology. Prehension processes would correspond to phenomenal counterparts to measurements and preparations.

The double slit experiment allows to illustrate the behavior of quantum systems during measurements. Suppose one has a board like a blackboard and we cut two vertical slits in it, a bit wider than a tennis ball. Now let us take one of those machines that shoot tennis balls and aim to the board. In classical mechanics the balls will either bounce on the board or make it through the slits. If the balls were covered in chalk, one would get a mark on the wall behind the board mirroring the slits as the balls make it through it and stain the wall. This would be the result when we are dealing with classical objects. If we now were to shrink the experiment to the micro size, where quantum mechanical effects are relevant, things would be different. We can construct analogous situations, for instance using crystals instead of the board with the slits and using the inter atoms spaces as slits and electrons instead of tennis balls and a photographic plate instead of the wall. We use a collimated beam of low intensity electrons so that we can assume that at each instant there is only one electron traveling from the source to the photographic plate. Each time an electron reaches the screen a small dot appear at some point of the plate.  After many electrons hit the plate, we would not see two vertical lines like in the macro world but an undulating range of intensity typical of an interference pattern. What is happening is that the electrons are behaving like waves when traveling from the source to the screen and as particles when their position on the plate is measured \cite{everyone}. The waves describe the behavior of an electron in a state represented by a wave packet that propagates through the slits and hits the plate. The interference pattern is typical of a wave; however, electrons behave as particles too. If one could do the experiment one electron at a time one would see them impinge on the photographic plate as a dot appearing at some position on it. But as more and more electrons pass, the dots will form an undulating pattern. The theory goes further in telling us that we cannot know through which of the slits the electron went through. If we try to measure it, for instance illuminating the slits to see through which one it went, the undulating pattern in the photographic plate changes. For the experiment to work we cannot know through which slit it passed. More generally, the quantum theory says that properties of the system do not take values until they are measured and therefore it does not make sense to ask through which slit an unobserved electron went. It is not that we do not know their values until they are measured, as is the case in classical mechanics, they simply cannot take values for the mathematical formalism of the theory to work properly \cite{everyone}. In fact, the interference figure appears even if the intensity of the beam is so low that there is only one electron hitting the screen at each instant. If the position of the electron took values as it traveled from the source to the plate, it would make sense to ask which slit each electron passed through, but if it could be said that each electron passed through one of the slits, there would not be any reason for the interference figure to be formed.

Phenomenologically, we consider that each time an electron travels from the source of collimated electrons to the photographic plate, a prehension occurs.  The object in this experiment is the photographic plate, and the proto-subject is the state of the electron coupled to the silver bromide molecules that cover the plate. The interaction of the electron with the photosensitive molecules induces a disposition to produce events consisting of the decomposition of the molecules at some position on the plate and the accumulation of silver atoms to form a dot at that point of the plate. The prehension with its corresponding affective tone would occur before the system chooses which event will appear on the plate. That is, the proto-subject chooses ---randomly--- the point where the electron will activate the decomposition of the silver bromide and the accumulation of silver atoms. A measurement is therefore a process of actualization of an event, like the formation of a dot at a point on the photographic plate. This interpretation is totally in line with what Whitehead established when he says that ``actualization is a selection among possibilities." \cite{whitehead} p. 159. The choice to which we have referred is physically preceded by a process known as environmental decoherence. In it, the state of the electron coupled to the plate loses quantum coherence because of its interaction with the plate environment. It behaves as a statistical mixture of electron states located in different regions of the plate and dots located in each of these regions. Decoherence occurs when a quantum system interacts with an environment with an enormous number of (microscopic) degrees of freedom. In this case, the state of the quantum system evolves into a state that ``almost” looks like the abrupt change one needs to postulate in measurements. After decoherence, the total state of the electron coupled to the photosensitive film of the plate takes a form that approximates the statistical mixture mentioned above \cite{zurek}. The affective tone in the provoked occasion mentioned by Whitehead would reach its climax at the instant of choosing between the different alternatives present in the state of the statistical mixture. The occasion of the experience would be, in Whitehead's terms, a subject with respect to the event chosen. The plate is the datum that provokes the origination of the prehension that occurs when the state of the electron coupled to the apparatus chooses one of the possible outcomes whose information is contained in the statistical mixture. 

Three phenomenal ingredients could be identified: a set of potential events, each with a certain disposition related to its probability of occurrence, a moment of climax in which a volitional act occurs ---a choice---, and the actualization of a qualitative feeling
associated with the chosen event. Notice that we are describing the prehension in terms of the ontology of states and events presented above: states correspond to disposition to act and events to phenomenal contents. As this ontology suggests, the prehension together with its qualitative feeling is not instantaneous, it accompanies the state, and during the measurement process it evolves when the object begins to interact with the 'object' composed by a measuring device, ---the photosensitive coating of silver bromide-- -, and its environment, ---the plate that supports this photosensitive film---. The process described here culminates with an event, and the phenomenal aspect related with its choice represents the proto-volitional act in which the dispositional state is updated by producing an event in the measuring device. The basic elements of a conscious volitional act should be of the same kind as the one previously discussed: an 'object' ---the brain----, a subject given the state of a particular set of neurons coupled to a quantum system in an entangled state ---for instance in the Fisher's model \cite{fisher} a multi-entangled set of Posner molecules---, where a prehension takes place activating an event: the firing of certain neurons that trigger an action. ``For an outside observer, the key characteristic of a voluntary act is that the voluntary act cannot be fully predicted from the preceding context... The behavior is not determined by external events, then the choices must be made 'from inside', endogenously" \cite{Frith} p. 290. The relevance of the climactic moment where a choice occurs and the state suffers an abrupt change leading to the occurrence of an event is directly related with the Whiteheadian notion of creativity. ``Creativity is the actualization of potentiality, and the actualization process is an occasion for experiencing" \cite{whiteheadaoi} p. 159. {\em Summarizing, a subject is the phenomenal side of a system in a certain state. Its life span is the interval between the event that prepares it and the one that it ends up producing in another system.}

Having analyzed the measurement of a state as a manifestation of a proto-volitional act, let us discuss the preparation of states and their relationship with proto perceptions. Usually preparing a specific state requires many compatible measurements, but an isolated measurement is enough to introduce a partial change of the state. For simplicity, we will analyze a case where the complete preparation process requires a single measurement. The phenomenological analysis of the general case can be carried out assuming that the different required measurements for the complete determination of the state are performed sequentially because they are compatible measurements. The following analysis applies to each measurement of the sequence. Recall that in a preparation, the events produced during the measurement that prepares the system put the subject in a new state. A state is born with the event that characterizes a phenomenal quality that may be considered as a proto-perception.

Let's take the process of preparing a particle with spin oriented in a certain $z$ direction pointing upward. The Stern–-Gerlach experiment exploits the magnetic properties of spinning particles. The main element is a magnet that creates a magnetic field between its poles that is more intense the closer one gets to the top pole. In the center the field points in the $z$ direction. That field exerts on a charged spinning particle like a silver atom injected into the region a force that would deflect the motion depending on the $z$ component of the spin. Atoms injected with spin up would move upwards and those with spin down downwards. Although their trajectories would not be precisely defined due to the uncertainty principle, they would lump together in two well differentiated regions. If one had done the experiment with classical particles, one would have got a continuous distribution depending on the initial orientation of the spin as they fly into the apparatus, because the spin component of a classical particle is not quantized. The experiment indicates that silver atoms have a quantized behavior with two possible values for its spin. The spins deviated to the upward region are called $ |z,up>$ and the ones deviated to the downward region $ |z,down>$.  An atom in an arbitrary state will be in a superposition of its two possible eigenstates $|\psi>= a |z,up>+b |z,down>$, with $a$ and $b$ complex numbers such that $|a|^2+|b|^2=1$, where $|a|^2$ is the probability of finding the atom in the upper region and $|b|^2$ the complementary probability of finding them downwards. Let us assume we initially have silver atoms with unknown spin ---that is, in a state $|\psi>$ with coeficients $a,b$ unknown--- and we want to use the Stern--Gerlach apparatus to prepare atoms whose spin is up and  whose state we denote by $|z,up>$. We proceed as follows. We consider an atom belonging to a collimated beam whose spin is unknown that is injected into the Stern--Gerlach device, and we put a detector in its downward region in such a way that if the atom goes there, it will be detected by  the observation of an event. Then the atom going through the apparatus either will be detected if it is deviated downwards or it will deviate upwards in the state $|z,up>$. Thus we end up with an atom prepared in the state  $|z,up>$ when no detection occurs. Notice that in quantum mechanics, even a negative measurement ---that is, if no particle is detected in the downward branch of the Stern--Gerlach device--- a change in the state occurs passing from $|\psi>$ to $|z,up>$. The change of state of the atom can be described again in terms of environmental decoherence: the incident wave packet representing the atom arriving to the Stern-Gerlach device interacts with the detector and its environment and its state initially in $|\psi>$ evolves into an approximate statistical mixture of states $|z,up>$ with probability $|a|^2$ and $|z,down>$ with complementary probability when a choice occurs leading to a change of the state of the atom now represented by $|z,up>$. Therefore, under the panprotopsychist hypothesis the atomic spin states play the role of subjects having an inner aspect characterized by a qualitative feeling or emotional tone that changes after the choice resulting from the presence of an ‘object’ ---the detector--- leading to new qualia or proto-perceptions in the subject. A prehension present in the incident atom perishes during the choice and a new prehension is established that will be present until the system goes through a new measurement process. Again, a proto subject’s physical counterpart is the spinning particle, a system in certain state, the preparation induces a prehension, in this case a proto-perception. 

Two subjects may combine in one if they interact and get entangled; in this case, the individuality of the parts is lost, and a new subject corresponding to the state of the two entangled particles appears. If the two spinning particles do not interact, the classical ontology would apply, and the final system will continue to be composed by the two subjects that in this case would preserve their individuality.  
Again, if we try to extrapolate from an elementary measurement to a mental process, the quantum system will be replaced by a multiparticle entangled state. Some of its components would interact with the neural system inducing a change of state in the quantum system whose phenomenal counterpart will be a new perception. As is now clear, being the total quantum system entangled, the states of its components lose individuality and therefore the only relevant subject is the corresponding to the total entangled system, as we have stressed when analyzing the palette and grain problems. Once one has identified the total entangled system as the subject of successive experiences, the subject summing problem admits the same solution as the other combination problems due to the quantum ontology. 

If what we identify as perception is a change of state during a measurement, there is an important part of the perceptual act that depends on complex functions of the nervous system that occur outside of consciousness and play the role of a measuring device. However, as we stressed in our analysis of the structuring of perceptions, the relationships between the physical qualities of the sensory input and the phenomenal content of the resulting perception should play a fundamental role in the development of the brain's perceptual systems that actively and preconsciously attempt to make sense of their inputs \cite{gregory} pp. 598–601.
In Husserl's own words: “experience is not an opening through which a world, existing prior to all experience, shines into a room of consciousness; it is not a mere taking of something alien to consciousness into consciousness... Experience is the performance in which for me, the experiencer, experienced being "is there", and is there as what it is, with the whole content and the mode of being that experience itself, by the performance going on in its intentionality, attributes to it. \cite{husserlformal} p. 232. 

As we have observed, prehensions are not instantaneous and have a certain temporal extension. Quantum ontology not only allows us to understand the combination problem, but also explains the temporality of phenomenal experience. In effect, we do not perceive time as a succession of instants. We do not perceive movement as a succession of positions; our now has a certain ``thickness". When listening to a melody, we do not hear an instantaneous note; we hear the entire sequence as if the present now retains the past. Husserl refers to retention as the process that preserves what just happened and protention to the anticipation of the future. We understand the language due to this process \cite{consciousness3}.

\section{Conclusions}

Physicalist visions, including Chalmers' natural supervenience,  are insufficient to understand the development of phenomenal capacities manifested in human consciousness, particularly the capacity to structure perceptual data. The highly rich and structured organization of our perceptions requires the influence of the phenomenal on our adaptive capabilities to allow the evolution of more primitive forms of experience manifested in animals, and to explain the very process of exploration and learning that children follow for the organization of perceptual data. If consciousness is merely epiphenomenal, how could we explain the development of the representational capacity involved in the perceptual abilities observed in birds, mammals, and humans? An epiphenomenal consciousness would also not account for several central manifestations in humanistic reflection, since ethical and aesthetic notions strongly depend on our freedom.

Scientific structuralism opens new perspectives that allow for a regularist form of panpsychism. That is, the notion that the regularities that physics identifies in its description of processes and phenomena are fulfilled without exception, without claiming that nature is interchangeable or exhausted by physical description. This is consistent with the structuralist conception of science shared by many philosophers and scientists such as Wittgenstein, van Fraassen, or Hawking that distinguishes between mathematical laws and the world they describe. Since regularism does not assume that the world is exhausted by the physical, a quantum probabilistic description is open to new causal powers such as those resulting from the phenomenal realm. If the phenomenal world has a degree of independence from the physical world and does not supervene on it, its most economical description is the panpsychist one. A mere epiphenomenic description would not be able to explain how our perceptual capacities develop. Consciousness must be effective.

Any object whose third-person empirical description is the quantum physical one could have a phenomenal intrinsic nature. The fundamental concepts of quantum mechanics according to its axioms are systems in certain states that produce events in their interaction with other systems. This ontology of states and events should be at the base of a panprotopsychic view founded on quantum mechanics. We have analyzed this possibility in a previous paper showing that it allows to solve several aspects of the combination problem \cite{consciousness3}, such as the palette and grain problems. Here we have extended this analysis to two of the oldest and most important objections to panpsychism: the structural and subject-summing combination problems related to the nature of perceptions and subjectivity, in particular of proto-subjectivity. 
The solution presents a remarkable degree of coherence and is based on the properties of entangled quantum states in a regularist context. Without regularism, important parts of the problem of structure and the subject-summing combination could not be addressed, much less understanding the role of freedom in art, ethics, and other forms of human behavior. Quantum panpsychism, due to its indeterminacy, allows for a regularist approach and could provide the possibility of addressing those questions.

\end{document}